\documentclass[11pt]{article}
\usepackage{moriond,epsfig}

\def\vev#1{\langle#1\rangle}

\def\be{\begin{equation}}
\def\ee{\end{equation}}
\def\bea{\begin{eqnarray}}
\def\eea{\end{eqnarray}}

\def\lesssim{\mathrel{\hbox{\rlap{\hbox{\lower4pt\hbox{$\sim$}}}\hbox{$<$}}}}
\def\gtrsim{\mathrel{\hbox{\rlap{\hbox{\lower4pt\hbox{$\sim$}}}\hbox{$>$}}}}

\begin{document}
\vspace*{4cm}
\title{EVOLUTION OF THE NON-LINEAR GALAXY BIAS UP TO REDSHIFT z=1.5}

\author{C. MARINONI and the VVDS consortium \footnote{O. Le F\`evre,
B. Meneux,
A. Iovino,
A. Pollo,
O. Ilbert,
G. Zamorani,
L. Guzzo,
A. Mazure,
R. Scaramella,
A. Cappi,
H.J. McCracken,
D. Bottini,
B. Garilli,
V. Le Brun,
D. Maccagni,
J.P. Picat,
M. Scodeggio,
L. Tresse,
G. Vettolani,
A. Zanichelli,
C. Adami,
S. Arnouts,
S. Bardelli,
J. Blaizot,
M. Bolzonella,
S. Charlot,
P. Ciliegi,
T. Contini,
S. Foucaud,
P. Franzetti,
I. Gavignaud,
B. Marano,
G. Mathez,
R. Merighi,
S. Paltani,
R. Pell\`o,
L. Pozzetti,
M. Radovich,
E. Zucca,
M. Bondi,
A. Bongiorno,
G. Busarello,
S. Colombi,
O. Cucciati,
Y. Mellier,
P. Merluzzi,
V. Ripepi,
D. Rizzo}}

\address{Centre de Physique Th\'eorique, UMR 6207, CNRS-Universit\'e de Provence, \\
Campus de Luminy, F-13288 Marseille, France}

\maketitle

\abstracts{
Deep redshift surveys of the universe provide the basic ingredients to
compute the  probability distribution function (PDF) of galaxy
fluctuations and to constrain its evolution with cosmic time. 
When this statistic is combined with analytical CDM predictions for the
PDF of mass, useful insights into the biasing function relating mass
and galaxy distributions can be obtained. In this paper, we 
focus on two issues: the shape of the biasing function and
its evolution with redshift. We constrain these quantities by using 
a preliminary sample of galaxies spectroscopically surveyed by the 
Vimos-VLT Deep Survey  in a deep cone $0.4<z<1.5$ covering 0.4 $\times$0.4 sq. deg.
We show that the ratio between the amplitude of galaxy fluctuations 
and the underlying mass fluctuations declines with cosmic time, and that its
evolution  rate  is a function of redshift: biasing evolution is marginal 
up to $z \sim 0.8$  and more pronounced  for $z > 0.8$.}

\section{Introduction}
If cold dark matter (CDM) exists, it is collisionless,  and interacts only gravitationally, 
then it is fair to say that we fully understand its spatial arrangement and clustering properties
on large scales. In particular we know how to characterize  the 
distribution properties  of the mass density contrast $\delta$, the fundamental
variable for Large Scale Structure studies.
Somewhat less clear, from a theoretical point of view,
is our quantitative understanding  of the complex mechanisms which, on various 
cosmological scales, 
regulate the formation and the evolution of luminous structures within the
underlying dark-matter distribution. 
In other terms, we still lack a 
reliable theoretical description of the distribution properties 
of the galaxy density contrast $\delta_g$. 
At present, only observations seem to offer the most promising way 
to constrain  the statistics of $\delta_g$. 

A related crucial issue is the understanding of
physical dependence of $\delta_g$ on the underlying dark matter 
fluctuation field $\delta$ on large scales. 
A comprehensive description of the "{\it biasing scheme}", {\it i.e.} 
of the functional relationship between galaxies and the underlying 
dark matter density fluctuations, lies at the hearth of all interpretations of 
Large Scale Strcture (LSS) theoretical models. Since structure formation models
all predicts the distribution of mass, the role of biasing is pivotal in mapping 
observations back onto the theoretical models.

While there is general observational consensus on the broad picture,
{\it i.e.} that primordial massive  galaxies form  inside dark matter halos
whose spatial distribution is  highly biased with 
respect to the underline mass distribution and that 
biasing must decrease as time goes by, the elucidation of
the finer details of this evolution as well as any meaningful
comparison with specific theoretical predictions is still far from
being secured.  Since clustering depends on morphology, color and
luminosity, and since most  high redshift samples have been
selected according to different colors or luminosity criteria, it is
not clear, for example, how the very different classes of objects
(Ly-break galaxies, extremely red objects or ultraluminous galaxies),
which populate different redshift intervals, can be considered a
uniform set of mass tracers across different cosmic epochs.
Furthermore, one must note that the biasing relation is likely
to be nontrivial, {\it i.e.} non-linear and scale dependent, especially at
high redshift \cite{sbd00,som01}.

Many approaches have been used to characterize the clustering of galaxies
and to understand its relation with the clustering of matter.
A complete specification of galaxy clustering may  be given by the full set 
of galaxy N-point correlation functions \cite{davpee}. This approach has been explored  
over the past decade as better and deeper redshift surveys have become available.

An alternative description may be given in terms of  the probability
distribution function (PDF) of a random field.
A PDF of the cosmological density fluctuations 
is the most fundamental statistic characterizing the large-scale structure of the 
universe. In principle, it encodes much of the information
contained within the full hierarchy of correlation functions, thus providing valuable 
information about gravitational evolution of density fluctuations.
While the shape of the PDF of mass fluctuations at any given cosmic
epoch is theoretically well constrained from CDM simulations, little is 
known about the observational PDF of the
general population of galaxies in the high redshift universe.
Even locally, this fundamental statistic has been often overlooked \cite{ost03}.

Using the first-epoch Vimos-VLT Deep Survey 
data \cite{lef05} we have derived the functional shape of the
PDF of galaxy overdensities  studying its evolution over the wide 
redshift range $0.4<z<1.5$. 
In particular we have shown \cite{mar02,mar05} how to derive
the  biasing relation $\delta_g=\delta_g(\delta)$ between galaxy and
mass overdensities from their respective PDFs $g(\delta_g)$ and $f(\delta)$.
As a matter of fact, assuming a one-to-one mapping between mass and galaxy overdensity 
fields, conservation of probability implies

\begin{equation}
\frac{d\delta_g(\delta)}{d\delta}=\frac{f(\delta)}{g(\delta_g)}.\label{probc}
\end{equation}

The advantage over other methods is that we can
explore the functional form of the relationship $\delta_g=b(z,\delta, R)
\delta$ over a wide range in mass density contrasts, redshift intervals
and smoothing scales R without specifying any {\it a-priori}
parametric functional form for the biasing function.

\section{The First-Epoch VVDS Redshift Sample}
The Vimos-VLT Deep Survey (VVDS) is a spectroscopic survey primarily designed for 
measuring more than 100,000 galaxy redshifts in the range $0<z<5$. 
The VVDS is an ambitious  observational program to study how different was the universe 
when half its current age. Fig 1 is an evocative picture which best tell us what VVDS
is: a genetic laboratory where we can hope to decode important information about the 
complex physical status of the present day universe by studying the Large Scale Structure 
in its embryonic form. 

\begin{figure}
\psfig{figure=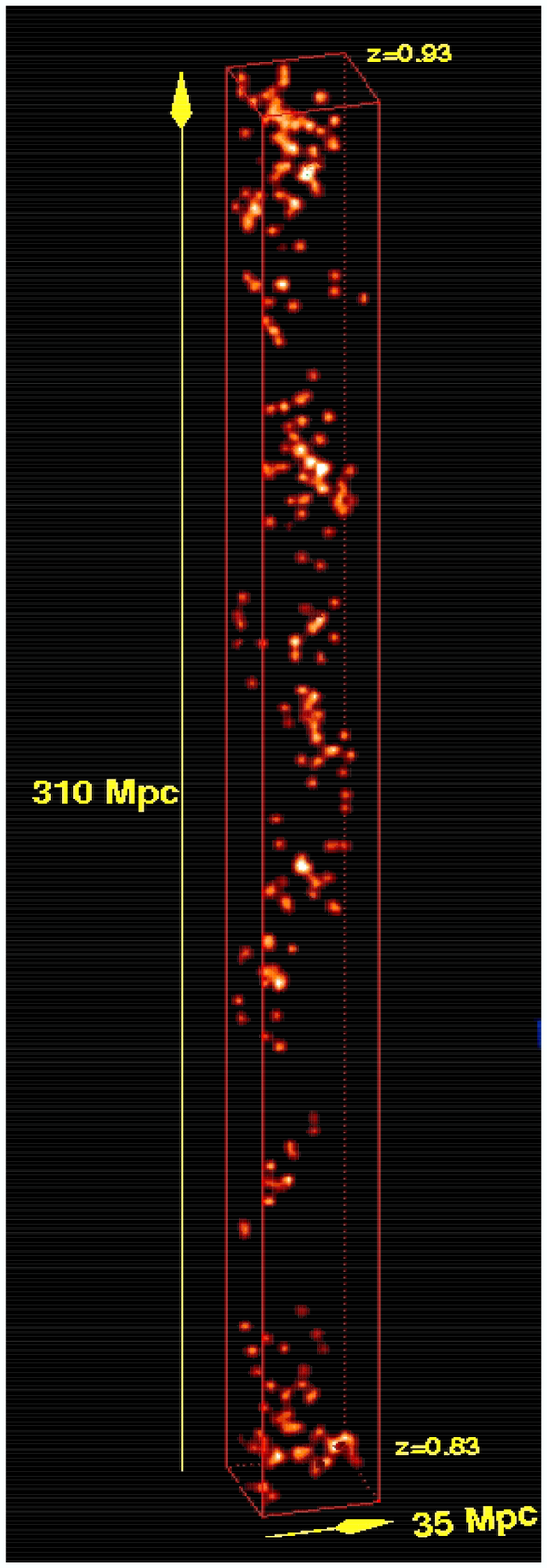,height=7.5in}
\caption{3D overdensity field traced by the galaxy distribution in the
VVDS deep  field. In order to obtain this geographical map of the distant 
regions of the Universe we have used 
data in the redshift interval $0.83<z<0.93$.
The galaxy distribution in this redshift range 
is continuously smoothed using a Top Hat window function with radius $R=2h^{-1}$Mpc.
The metric has been  computed assuming a $\Lambda$CDM
cosmology and the correct axis ratio between transversal and radial
dimensions has been preserved.  The approximate transverse
and radial  dimensions of the volume are shown in the figure.
\label{unof}}
\end{figure}

The  VVDS studies the evolutionary sequence of galaxies, 
clusters and AGNs with a double observational strategy : {\it a)} 
with  a {\it wide} survey which covers 16 deg$^2$ down to the  
limiting magnitude $I_{AB}=22.5$ and {\it b)} with a {\it deep} survey covering about 1.3 deg$^2$
down to $I_{AB}=24$.
The strength of the VVDS, compared to other currently undergoing deep surveys 
of the universe, is that  it has
been conceived as a purely flux-limited survey, {\it i.e.}  no target
pre-selection according to colors, morphology  or compactness is implemented.
As a consequence its selection function is simple to understand and it allows a direct 
and easier comparison between the high-z and low-z samples of galaxies.

The analysis presented in this paper is based on data collected in  the deep 
VVDS-02h field. In this field ($0.4 \times 0.4$ sq. deg.)
VIMOS observations have been performed using 1" wide
slits and the LRRed grism which covers the spectral range
$550<\lambda($nm$)<940$ with an effective spectral
resolution $R \sim 227$ at $\lambda=750$nm.  The accuracy in redshift measurements is
$\sim$275 km/s.  Details on 
observations and data reduction are given elsewhere \cite{lef05,lef04}.

For the purposes of this study we have defined a VVDS sub-sample  
with galaxies having redshift  $z<1.5$. 
Even if we measure
redshifts up to $z\sim$5, 
the conservative redshift limits 
bracket the range where we can sample in a denser way
the underlying galaxy distribution and, thus,
minimize biases in the reconstruction of the density field.
This  subsample contains 3448 galaxies with secure redshift 
and the redshift sampling rate is $\sim 30\%$
{\it i.e.}, down to $I_{AB}=24$, we measure redshifts for nearly  one over three galaxies.

It is worth to emphasize that the  VVDS
is probing the high redshift domain at I$\leq$24 in the VVDS-02h-4 field
with the same sampling  rate of pioneer surveys of the local Universe  such as 
the CFA (at z$\sim$0) and, more recently, the 2dFGRS \cite{coll01} (at z$\sim$0.1).

\section{Method}

It seems now well established \cite{ver02,lah02} that, in the local universe, 
the distribution of baryonic matter does form a faithful representation 
of the spatial properties of the dominant species of matter 
({\it i.e.} collisionless weakly interacting dark matter). On the contrary, very little 
is known about the rate of evolution as a function of redshift 
of the biasing relationship. Even less it is known about possible deviations, on large scales,  
from the simple linear parameterization which is almost universally 
adopted in order to describe of the biasing function on large scales (i.e. $R>5 h^{-1}$Mpc).

By  using  $\delta_g (\delta)=b(z,\delta, R) \delta$, 
into eq. \ref{probc},
we have derived   the redshift-, density-, and
scale-dependent biasing function $b(z, \delta, R)$ between galaxy and
matter fluctuations in a $\Lambda$CDM universe as the solution of the following
differential equation

\begin{equation} \left\{ \begin{array}{l} \delta_g(-1)=-1 \\ \\
b^{'}(\delta)\delta+b(\delta)=f(\delta) g(\delta_g)^{-1} \end{array}
\right.  \label{de} \end{equation}

\noindent where the prime denotes the derivative with respect to
$\delta$, $f(\delta)$ and $g(\delta_g)$ are the PDF of mass and galaxy
fluctuations respectively, and the initial condition has been
physically specified by requiring that galaxies cannot form where
there is no mass.

With this computational approach, we loose information on a possible
stochasticity characterizing the biasing function.
The advantage is that we can provide a
preliminary measure, on some characteristic scales R, of the {\it
local}, {\it non-linear}, {\it deterministic} biasing function
over the continuous redshift interval $0.4<z<1.5$.

Note that in our computational scheme, we explicitly assume that 
the mass PDF is described, to a good approximation, by a  log-normal distribution \cite{col91}

\begin{equation}
f(\delta)=\frac{(2 \pi \omega^2)^{-1/2.}}{1+\delta} \exp \Big\{ -\frac{
[\ln(1+\delta) +\omega^2/2]^2}{2\,\omega^2} \Big\} \label{teopdf}
\end{equation}

\noindent  characterized by a single parameter ($\omega$) that 
is related to the variance of the $\delta$-field as

\begin{equation} \omega^2=\ln [1+\vev{\delta^2}]
\label{omega}
\end{equation}

Particular attention has been paid to
devise an optimal strategy so that the
comparison of the PDFs of mass and galaxies can be carried out in
an objective and accurate way. 
First we have tested 
the statistical reliability of the observationally inferred PDF of VVDS
galaxy fluctuations. 
By applying the VVDS observational selection functions to GALICS semi-analytical
galaxy simulations \cite{hat} we have  explored the region of the parameter space
   where the PDF of VVDS-like  densities traces in a
   statistically unbiased way the parent underlying PDF of the real
   distribution of galaxy overdensities.
We have shown \cite{mar05} that the observed PDF
of  galaxy density contrasts  is an   unbiased  tracer of the underlying
distribution   up   to redshift  $z=1.5$,   on  scales  $R\gtrsim 8 h^{-1}$ Mpc.

Second, we have corrected the log-normal approximation, which describes
the mass density PDF, in order to take into account redshift
distortions induced by galaxy peculiar velocities
at early cosmic epochs, {\it i.e.} when the mapping between redshifts
and comoving positions is not linear. In this way, the theoretically
predicted mass PDF  can be directly compared to the corresponding 
observational quantity (galaxy PDF) directly  in redshift space.

Finally, we note that, in pursuing our approach, we have  assumed that 
the current theoretical
understanding of how clustering of DM proceeds via gravitational
instability in the expanding universe is well developed, {\it i.e.} the PDF
$f(\delta)$ of mass fluctuations can be safely derived via
analytical models or N-body simulations. 
In particular, in what follows, we will consider a $\Lambda$CDM background 
mass distribution locally normalized to $\sigma_8(z=0)=0.9$.

\section{Results}

\begin{figure}
\begin{center}
\psfig{figure=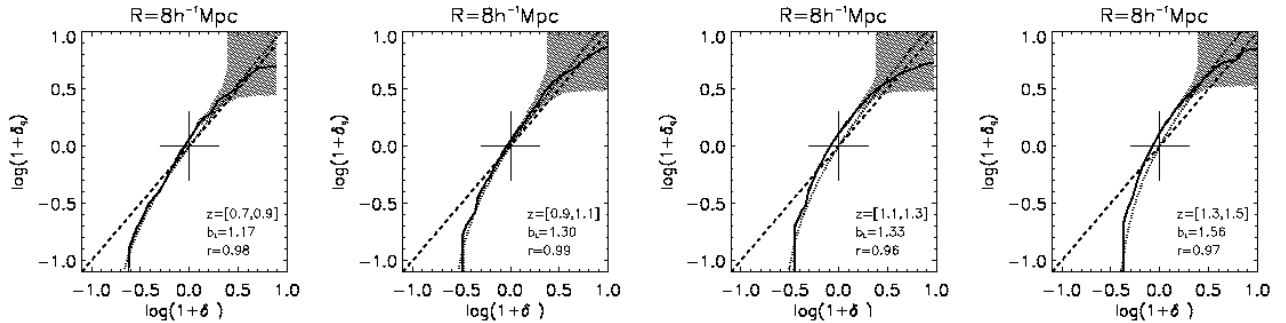,height=1.75in}
\caption{The observed biasing
function (solid-line) recovered for the
density field smoothed on scales $R=8h^{-1}$Mpc 
and for different redshift bins (from left to right)
in the volume-limited VVDS sample ($\mathcal{M}^{c}_B=-20+5\log h$).
The dotted line represents the linear biasing model $\delta_g=b_L \delta$ while
the no-bias case ($b_L =1$) is shown with a dashed line.
The central cross is for reference and represents the $\delta_g=\delta=0$ case.
The shaded area represents $1\sigma$ errors  in the derived biasing
function but do not include uncertainties due to cosmic variance
}
\end{center}
\end{figure}

We have obtained the biasing function $b(\delta)$ by numerically
integrating the differential equation (\ref{de}) {\it i.e.} 
without {\it a-priori} parameterizing the form of the biasing function.
We have solved eq. \ref{de} {\it i)} in different
redshift intervals,  {\it ii)} using matter and galaxy PDFs
obtained by smoothing the density fields on $R=8h^{-1}$Mpc
and {\it iii}) using a volume-limited sub-sample of galaxies with absolute 
magnitude $\mathcal{M}_{B}<-20+5\log h$.

Results are presented in Fig. 2 where we show that the biasing function has 
in general a
non trivial shape and where we  trace its evolution across different cosmic epochs.
The  main conclusions inferred by analyzing the solution of eq. \ref{probc} 
are presented and discussed  in detail by Marinoni et al. 2005 \cite{mar05}.
Here we briefly summarize  our main  findings.

{\it i}) In general, the linear approximation offers in general 
a poor description of the richness 
of details encoded in the biasing function, {\it i.e.} the linear biasing function 
(dotted line in Fig.2)  poorly describes
the observed scaling of the $\delta_g$ vs. $\delta$ 
relation (solid line). 
Non-linear effects in the biasing relation  are detected  at a level
of $\lesssim 10\%$. In particular the ratio 
between the quadratic and linear term of the biasing
expansion \begin{equation}
\delta_g=\sum_{k=0}^{2}\frac{b_k}{k!}\delta^k. \label{tayl}
\end{equation}
is nearly constant in the redshift range $0.7<z<1.5$ and 
different from zero
at a confidence level greater than 3$\sigma$
(we find $b_2/b_1\sim-0.15\pm0.04$ for $R=8h^{-1}$Mpc).
This result confirms a  general prediction
of CDM-based hierarchical models of galaxy formation \cite{som01}.
However, such non-linear distortions of the biasing function are  not observed locally
in the 2dFGRS sample \cite{ver02}. The fact that 
the  $b_2/b_1$ ratio is not only different from zero but also 
remains constant  over all the redshift intervals investigated in the 
range $0.7<z<1.5$  is even more difficult 
to reconcile  with the "null" result of local measurements.

\begin{figure}
\begin{center}
\psfig{figure=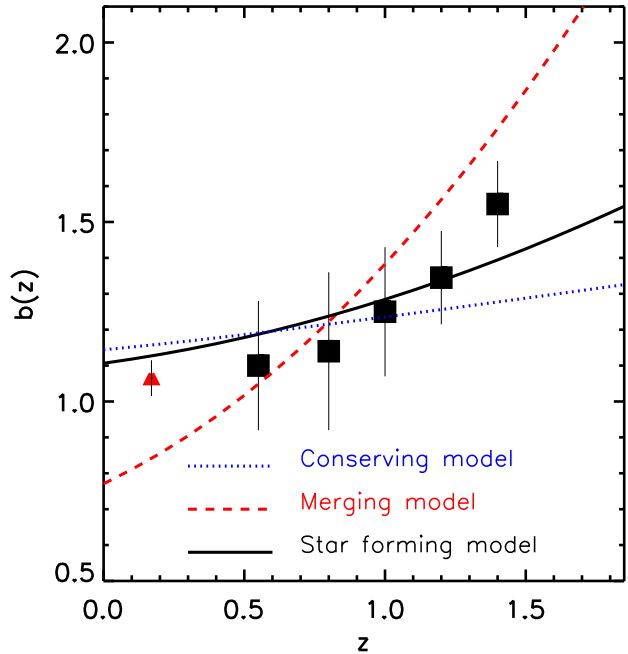,height=3.5in}
\caption{ The redshift evolution of the
linear biasing parameter  on a $8h^{-1}$Mpc scale  for
the volume-limited ($\mathcal{M}^{c}_{B}<$-20+5$\log$h) VVDS subsample
(filled squares) is shown. 
The triangle represents the $z \sim 0$ bias
inferred  for 2dFGRS galaxies having median $L/L^*\sim 2$
({\it i.e.} the median luminosity of the volume-limited VVDS sample).
Our measurements are also compared to
various theoretical models of biasing evolution.
The dotted line indicates the conserving model (Fry 96), the solid and dashed lines
represent the star forming (Mo \& White 1996)  and merging (Blanton et al. 2000) 
models.}
\end{center}
\end{figure}

{\it ii}) Non-linear effects bend the biasing function 
in such a way that the $\delta_g$ vs. $\delta$ relation is steeper
in underdense regions (the local slope is  $b(\delta)>1)$ 
than in overdense ones. In particular, Fig. 2 shows that 
below some finite mass density threshold  the formation efficiency 
of galaxies brighter than $\mathcal{M}_{B}<-20+5\log h$ drops to zero. 
Moreover the mass-density threshold below which the formation of
   bright galaxies ($\mathcal{M}_{B}<-20+5\log h$) seems to be inhibited 
decreases as a function of time. This trend suggests that galaxies
   of a given luminosity were tracing systematically higher mass overdensities 
in the early Universe, and that 
   as time progresses, galaxy formation begins to take place also in lower density peaks.
Stated differently, the assembling of luminous galaxies in low density
regions becomes more efficient as time goes by.

{\it iii}) We do not observe the imprints of scale-dependency in the biasing function
a behavior in agreement with  results derived from local surveys \cite{ver02} at z$\sim$ 0.

{\it iv}) By representing the biasing function in linear approximation \cite{ld99}, {\it i.e.} by 
compressing all the information contained in the $\delta_g$ vs. $\delta$ 
relation into a single scalar parameter $b_L$ using the equation
\begin{equation}
b_L^2 \equiv \frac{\vev{b^2(\delta)\, \delta^2}}{\vev{\delta^2}}
\end{equation}
\noindent  it is easier to compare our results with those of other authors 
(who often "{\it a-priori}" adopt a linear parameterization for describing galaxy bias) 
and also with predictions of theoretical models.
With this approximation, 
we have found that the linear biasing parameter evolves with cosmic time.
It appears that we live in a special epoch in which the galaxy distribution
traces the underlying mass distribution on large scales ($b_L \sim 1$), while,
in the past, the two fields were progressively dissimilar
and the relative  biasing systematically higher.
The difference between the value of $b_L$ at redshift z$\sim$1.5 and z$\sim$ 0
for a population of galaxies with luminosity $\mathcal{M}_{B}<-20+5\log h$
is significant  at a confidence level greater than 3$\sigma$
($\Delta b_L \sim 0.5\pm0.14$). In this interval,  the essential
characteristics of the  time evolution of the linear
bias are well described in terms of  the phenomenological  relationship
$b_L=1+(0.03\pm0.01)(1+z)^{3.3+0.6}$.
Over the redshift baseline investigated, the rate of biasing evolution is a function of redshift: 
z$\sim$0.8  is the  characteristic redshift which marks the transition
from a ``minimum-evolution'' late epoch to an early period 
where the biasing evolution for a population of $\mathcal{M}_{B}<-20+5\log h$ galaxies 
is substantial ($\sim 33\%$ between redshift 0.8 and 1.4).

{\it v}) Even at past epochs, brighter galaxies were  more strongly 
biased than  less  luminous ones. Moreover  the dependence of  biasing on luminosity at
$z\sim$0.8 is  in good  agreement with what  is observed  in  the local
universe \cite{nor01}. In other terms even at high redshift, luminous galaxies avoid 
mass underdense regions while fainter ones are found also in low density environments.

{\it vi}) By comparing  our results to predictions of theoretical models 
for the  biasing evolution,  we have shown that   the galaxy {\it conserving} 
model \cite{fry96} and halo {\it merging} \cite{mow96} model offer a poor 
description of our data (See Fig 3). This result could  suggest that the gravitational 
debiasing or the hierarchical merging of halos alone 
may not be the only physical mechanisms driving
the evolution of galaxy biasing across cosmic epochs. At variance with these results, the 
{\it star forming} model \cite{bla} seems to describe better the observed  redshift evolution of biasing. However the conclusion we can draw is that these three different models which are 
based on  simplifying  hypothesis, if considered alone,  are far from being realistic.
we need a more  complex modeling of biasing evolution.
Our analysis  seems to suggest the apparent need of more complex biasing mechanisms
to explain the observed biasing evolution.

{\it vii}) The red sample is  systematically 
a more biased tracer of mass  than the blue one in every redshift interval investigated, but 
the relative biasing between the two populations  is nearly   constant in the
redshift range 0.7$<z<$1.5 ($b^{red}/b^{blue}\sim 1.4\pm0.1$), and  comparable with local estimates (see also Meneux et al. 2006 \cite{men06}).
Moreover, we have found that the  bright red subsample is biased with respect to
the general red population in the same way as the bright sample of blue objects 
is biased with respect to the global blue population thus indicating, 
that  biasing as a function of luminosity might be,  at first order, 
independent of colors. 

{\it viii}) Due to the large errorbars which still affect our results, the bias of our sample 
of bright and moderately red objects at z$\sim$1 is not statistically dissimilar from that expected for EROS of similar 
luminosity, even if the EROS biasing appears to be systematically larger.

One key aspect of this study is the measure of evolution in the distribution
properties of galaxy overdensities from a continuous volume sampled with the
same selection function over a wide redshift baseline.
In a different paper (Marinoni {\it et al.} in prep.)
we discuss the cosmological implications of our results namely by 
testing the standard assumption that the structure we see today are
the results of the  gravitational amplification and collapse of small
primordial matter density fluctuations.
As the volume sampled is still limited, errors on the analysis presented in this paper
are dominated by cosmic variance.
The technique presented here will be applied to a larger sample as the VVDS
observational program  progresses.

\section*{Acknowledgments}
This research has been developed within the framework of the VVDS consortium
and it has been partially supported by the CNRS-INSU and its Programme
National de Cosmologie (France), and by the Italian Ministry (MIUR) grants
COFIN2000 and COFIN2003.
The VLT-VIMOS observations have been carried out on guaranteed
time (GTO) allocated by the European Southern Observatory (ESO) to the
VIRMOS consortium, under a contractual agreement between
the Centre National de la Recherche Scientifique of France, heading
a consortium of French and Italian institutes, and ESO, to design, manufacture
and test the VIMOS instrument.


\section*{References}
\begin{thebibliography}{99}
\bibitem{sbd00}  Sigad, Y., Branchini, E., Dekel, A., 2000, ApJ, 540, 62
\bibitem{som01}  Somerville, R. S., Lemson, G., Sigad, Y., Dekel, A., Kauffmann, G., White, S. D. M., 2001, MNRAS 320, 289.
\bibitem{davpee} Davis, M. \& Peebles, P. J. E. 1977 ApJS, 34, 425
\bibitem{ost03}  Ostriker, J. P., Nagamine, K., Cen , R., Fukugita, M., 2003, ApJ, 597, 1
\bibitem{lef05}  Le F\`evre, O., et al. 2005, A\&A, 439, 845
\bibitem{mar02}  Marinoni, C. \& Hudson, M., 2002, ApJ, 569, 101
\bibitem{mar05}  Marinoni, C., et al. 2005, A\&A, 442, 801
\bibitem{lef04}  Le F\`evre, O., et al. 2004, A\&A, 428, 1043 
\bibitem{coll01} Colless, M. M., 2001, MNRAS, 328, 1039
\bibitem{ver02}  Verde, L., et al., 2002, MNRAS, 335, 432
\bibitem{lah02}  Lahav, O., et al. 2002, MNRAS, 333, 961
\bibitem{hat}    Hatton,S., Deviendrt, J. E. G., Ninin, S., Bouchet, F. R., Guiderdoni, B., \& Vibert, D.  2003, MNRAS, 373, 75
\bibitem{col91}  Coles, P., \& Jones, B. 1991, MNRAS, 248, 1
\bibitem{ld99}   Dekel, A., \& Lahav, O.  1999, ApJ, 520, 24
\bibitem{nor01}  Norberg, P.  et al. 2001, MNRAS, 328, 64
\bibitem{fry96}  Fry, J. N. 1996, ApJ, 461, L65
\bibitem{mow96}  Mo, H., White, S. D. M., 1996, MNRAS, 282, 347
\bibitem{bla}    Blanton, M., Cen, R., Ostriker, J. P., Strauss, M. A., Tegmark, M., 2000, ApJ, 531, 1
\bibitem{men06}    Meneux, B., et al. 2006, A\&A, 452, 387 

\end{thebibliography}
\end{document}